\documentstyle[preprint,aps]{revtex} \par \begin{document} \par {\tighten
\preprint{\vbox{ \hbox{FERMILAB--PUB--96/448--T} \hbox{BNL- } \hbox{JLAB-TH-96-
} }} \title{\bf Enhanced CP Violation with $B\to K D^0 (\overline D^0)$ Modes
and Extraction of the CKM Angle $\gamma$} \author{David Atwood$^a$, Isard
Dunietz$^b$ and Amarjit Soni$^c$} \address{\it a) Thomas Jefferson National
Accelerator Facility, Newport News, VA 23606} \address{\it b) Fermi National
Accelerator Laboratory, P.O. Box 500, Batavia, IL 60510} \address{\it
c)Brookhaven National Laboratory, Upton, NY 11973} \par \bigskip \maketitle
\par \begin{abstract} The Gronau-London-Wyler (GLW) method extracts the CKM
angle $\gamma$ by measuring $B^\pm$ decay rates involving $D^0/\overline D^0$
mesons. Since that method necessitates the interference between two amplitudes
that are significantly different in magnitude, the resulting asymmetries tend
to be small. CP violation can be greatly enhanced for decays to final states
that are common to both $D^0$ and $\overline D^0$ and that are not CP
eigenstates. In particular, large asymmetries are possible for final states $f$
such that $D^0\to f$ is doubly Cabibbo suppressed while $\overline D^0\to f$ is
Cabibbo allowed. The measurement of interference effects in two such modes
allows the extraction of $\gamma$ without prior knowledge of $Br(B^-\to K^-
\overline D^0)$, which may be difficult to determine due to backgrounds.
\end{abstract} } \par \bigskip \par \bigskip \par One striking implication of
the standard model with three families is that it can accommodate CP violation 
via the Kobayashi-Maskawa mechanism~\cite{ckmref}. Intense experimental efforts
are now underway in $B$-physics to test the standard model in this regard
through measurements of the unitarity triangle~\cite{quinpdb}. For this program
to succeed it is of crucial importance to be able to deduce each of the angles
of this triangle from experiment. In this paper we will focus our attention to
one of the three angles, namely $\gamma$. \par We recall that in the standard
model, $b\to c \overline u s$ and $b\to \overline c u s$ transitions have a
relative Cabibbo-Kobayashi-Maskawa (CKM) phase $\gamma$. In order to measure CP
violation due to this phase, a means must be found to have these seemingly
distinct final states interfere. A mechanism whereby this is possible has been
proposed and extensively studied
\cite{bigis,gronaul,gronauw,d0ks,baryon,feasibility}. The basic idea is that if
the $\overline u c$ ($\overline c u$) hadronize into a single $D^0$ ($\overline
D^0$) meson, which is subsequently seen as a CP eigenstate (e.g. $K_S\pi^0$) or
$K_S+n\pi$, then both processes lead to a common final state.  These two
channels can thus interfere quantum mechanically giving rise to, in particular,
CP violating effects~\cite{bigis}. \par The Gronau-London-Wyler (GLW)
method~\cite{gronaul,gronauw,d0ks,baryon,feasibility} extracts the CKM angle
$\gamma$ from measurements of the branching ratios of the six processes,
$B^-\to K^- \overline D^0,\ K^- D^0,\  K^- D^0_{CP}$ and their CP-conjugate
partners. Here $D^0_{CP}$ denotes that the $D^0$ or the $\overline D^0$ is seen
in a CP eigenstate. The two interfering amplitudes have a CP violating phase
$\gamma$ between the $D^0$ and the $\overline D^0$ paths leading to the common
final state. The manifestation of CP violation also requires a CP even phase
difference. This will generally be present due to final state interactions
although it is not known how to calculate it reliably. However, even if this
strong phase difference is small, information about $\gamma$ may still be
extracted from CP even interference effects. \par The use of $D^0$ and
$\overline D^0$ decays to common states that are {\it not CP eigenstates} was
proposed several years ago~\cite{baryon}. In this Letter we wish to point out
that among this category, $D^0$ decays which are doubly Cabibbo suppressed lead
to CP violating effects that may be greatly enhanced. In addition, a number of
potential experimental difficulties with the GLW method may be reduced or
overcome. \par The primary problem with respect to the GLW method is the fact
that CP violating asymmetries tend to be small since $B^-\to K^- \overline D^0$
is color suppressed whereas $B^-\to K^- D^0$ is color allowed. Moreover, when
the appropriate CKM factors are taken into account, the former amplitude is
typically an order of magnitude smaller than the latter. In the GLW method the
interference effects are therefore limited to $O(10\%)$, which indicates the
maximum possible size for CP violation via this method. To overcome this we
choose instead $D^0$-modes, $f$, that are not CP-eigenstates. Especially
appealing are modes $f$ such that $D^0\to f$ is doubly Cabibbo suppressed while
$\overline D^0\to f$ is Cabibbo allowed (e.g. $f=K^+\pi^-,\;K \pi\pi,\;$ etc.).
As a result, the two interfering amplitudes become comparable. Numerically, the
ratio between these two amplitudes is crudely given by~\cite{suppressed}:
\begin{eqnarray} \label{ampratio} \left |\frac{{\cal M}(B^- \to K^- D^0 [\to
f])} {{\cal M}(B^- \to K^- \overline D^0 [\to f])}\right |^2 & \approx & \left
|\frac{V_{cb} V^*_{us}}{V_{ub} V^*_{cs}}\right |^2 \;\;\left
|\frac{a_1}{a_2}\right |^2 \;\;\frac{Br(D^0 \to f)}{Br(\overline D^0 \to f)}
\approx \\ \nonumber & \approx & \left |\frac{0.22}{0.08}\right |^2 \;\;\left
|\frac{1}{0.26}\right |^2 \;0.0077 \sim 1 \;, \nonumber \end{eqnarray} where
${\cal M}$ denotes the amplitude for the given process. Here the
color-suppressed amplitude $(\sim a_2$) is reduced with respect to the
color-allowed one $(\sim a_1$) by the factor suggested in~\cite{browder}:
\begin{eqnarray} |{a_2}/{a_1} | \approx 0.26, \nonumber \end{eqnarray} and the
ratio of CKM elements $|V_{ub}/V_{cb}|\approx 0.08$ was used. \par While a
naive estimate for the ratio of twice Cabibbo suppressed to Cabibbo-allowed
branching ratio is \begin{eqnarray} \frac{Br(D^0 \to f)}{Br(\overline D^0 \to
f)} \approx \left |\frac{V_{cd} V_{us}}{V_{cs}V_{ud}}\right |^2 \approx
\lambda^4 \;, \end{eqnarray} form-factor and decay constant ratios may increase
it somewhat. Such a ratio has been observed by CLEO \cite{yamamoto}
\begin{eqnarray} \label{dcsexp} \frac{Br(D^0 \to K^+ \pi^- )}{Br(\overline D^0
\to K^+\pi^- )} = 0.0077 \pm 0.0025 \pm 0.0025\;, \nonumber \end{eqnarray}
whose central value was used in Eq.~(\ref{ampratio}) for the generic ratio.
\par The balancing of the amplitudes  illustrated in Eq.~(\ref{ampratio})
suggests that CP violating effects in the interference of two amplitudes of
this type can be large. Let us define, for a general final state $f$, the CP
violating partial rate asymmetry: \begin{eqnarray} \label{cpasym}
A(K,f)\equiv\frac{Br(B^-\to K^- [f]) - Br(B^+\to K^+ [\overline f])}{Br(B^-\to
K^-[f]) + Br(B^+\to K^+ [\overline f])}\; \nonumber \end{eqnarray} where the
square bracket denotes that the bracketed mode originates from a $D^0/\overline
D^0$ decay. Based on the above argument potentially the largest CP violating
asymmetry $A(K,f)$ in $B^\pm$ decays involving $D^0-\overline D^0$ interference
occurs when $f$ is a doubly Cabibbo suppressed decay mode of the $D^0$. \par In
the  GLW method where $f$ is a CP eigenstate, the strong phase difference
between $D^0\to f$ and $\overline D^0\to f$: \begin{eqnarray} \delta_f={\rm
arg}({\cal M}(D^0\to f){\cal M}(\overline D^0\to f)^*) \nonumber \end{eqnarray}
is to an excellent approximation $0\;$mod$\;\pi$~\cite{foot}. Therefore the
total strong phase difference involved is that of the initial $B$ decay,
$\zeta_K\;$mod$\;\pi$, where $\zeta_K$ is given by: \begin{eqnarray}
\zeta_K={1\over 2} {\rm arg} \left [ {\cal M}(B^-\to K^- D^0){\cal M}(B^-\to
K^- \overline D^0)^* {\cal M}(B^+\to K^+ D^0)^*{\cal M}(B^+\to K^+ \overline
D^0) \right]\;. \nonumber \end{eqnarray} Since $A(K,f)\propto
\sin(\zeta_K+\delta_f)=\pm\sin(\zeta_K)$, if $\zeta_K$ should happen to be
small the GLW method will produce only a small CP violating signal. In
contrast, for non-CP eigenstates $f$, $\delta_f$ may assume different values,
some of which could be large. Indeed some experimental evidence suggests that
final state interaction effects in such $D^0$ decays can be
appreciable~\cite{lebrun}. Since several such modes are experimentally
feasible, for instance $f=K^+\pi^-$, $K^+\rho^-$, $K^+a_1^-$, $K^{*+}\pi^-$,
$K\pi\pi$, etc., it is likely that for at least some of these
$\sin(\zeta_K+\delta_f)$ will be large leading to a large asymmetry $A(K,f)$.
\par Another potential problem that arises with the GLW method is that to
reconstruct $\gamma$ it is necessary to know separately the branching ratios
$Br(B^-\to K^- D^0)$ and $Br(B^-\to K^- \overline D^0)$. While $Br(B^-\to K^-
D^0)\sim O(10^{-4})$ can be measured via conventional methods, $Br(B^-\to K^-
\overline D^0)\sim O(10^{-6})$ suffers from some serious experimental
difficulties. \par First, if $Br(B^-\to K^- \overline D^0)$ is measured through
the use of hadronic decays of the $\overline D^0$ (e.g. $\overline D^0\to
K^+\pi^-$) then, as Eq.~(\ref{ampratio}) demonstrates, interference effects of
$O(1)$ with the $D^0$ channel (e.g. $B^-\to K^- D^0 [\to K^+\pi^-]$ ) will be
present. Clearly then, the $\overline D^0$ must be tagged with a decay that is
distinct from any decay of the $D^0$, for instance the semileptonic decay
$\overline D^0\to l^- \overline \nu_l X_{\overline s}$. This mode, however is
subject to daunting backgrounds, such as $B^-\to l^-\overline\nu_l X_c$ which
is $O(10^6)$ times larger. Such backgrounds may be difficult to
overcome~\cite{d0bartag}. \par In our technique, the possibility of having a
variety of strong phases allows for several methods for the extraction of
$\gamma$~\cite{ads2}. For brevity, we will mention only two in this Letter. We
assume here all relevant branching ratios for $D^0$ decays are known. \par In
method (1) we assume that $Br(B^-\to K^- D^0)$ is known but not $Br(B^- \to
K^-\overline D^0)$. We also require the experimental determination of BR's for
at least two distinct final states $f_1$ and $f_2$ (where at least one of
$f_1$, $f_2$ is not a CP eigenstate): \begin{eqnarray} Br\left(B^- \to K^-
\left[f_i \right]\right) \;, \;\;Br\left(B^+\to K^+ \left[\overline
f_i\right]\right)\;, \ \ \ {\rm for}\ \ \ \ i=1,2\; . \nonumber \end{eqnarray}
This information suffices to extract $\gamma$, $Br(B^-\to K^- \overline D^0)$,
and the two relevant strong phase differences up to some discrete ambiguity.
\par To see how this works, let us define the quantities: \begin{eqnarray} &
a(K)=Br(B^-\to K^- D^0) \ \ \ b(K)=Br(B^-\to K^- \overline D^0) \ \ \
c(f_i)=Br(D^0\to f_i) & \nonumber\\ & c(\overline f_i)=Br(D^0\to \overline f_i)
\ \ \ d(K,f_i)=Br(B^-\to K^- [f_i]) \ \ \ \overline d(K,f_i)=Br(B^+\to K^+
[\overline f_i]) & \nonumber \end{eqnarray} where $i=1,2$. In this case,
therefore, we know the quantities $a(K)$, $c(f_i)$, $c(\overline f_i)$,
$d(K,f_i)$,  $\overline d(K,f_i)$ but not $b(K)$. \par The expressions for
$d(K,f_i)$,  $\overline d(K,f_i)$ in terms of the strong phases and $\gamma$
gives us four equations: \begin{eqnarray} d(K,f_i)&=&
a(K)c(f_i)+b(K)c(\overline f_i) +2\sqrt{a(K)b(K)c(f_i)c(\overline
f_i)}\cos(\xi^K_{f_i}+\gamma) \nonumber\\ \overline d(K,f_i)&=&
a(K)c(f_i)+b(K)c(\overline f_i) +2\sqrt{a(K)b(K)c(f_i)c(\overline
f_i)}\cos(\xi^K_{f_i}-\gamma) \label{eqnd} \end{eqnarray} where
$\xi^K_{f_i}=\zeta_K+\delta_{f_i}$. These four equations contain the four
unknowns $\{\xi^K_{f_1}$, $\xi^K_{f_2}$, $b(K)$, $\gamma\}$ which therefore can
be determined up to discrete ambiguities. Adding additional modes will, in
general, reduce the ambiguity to an overall two-fold one in the sign of all the
phases. \par This method also illustrates the importance of $D$ decay studies
in interpreting such CP violation in $B$ decays. The strong phases
$\xi^K_{f_i}$ relevant to eq.~(\ref{eqnd}) are related to the $D$ decay phase
shifts $\delta_{f_i}$ via \begin{eqnarray} \xi^K_{f_1}-\xi^K_{f_2}
=\delta_{f_1}-\delta_{f_2} \label{xidelta} \end{eqnarray} Since the separate
phase shifts $\delta_{f_i}$ on the right hand side of (\ref{xidelta}) may be
determined from data at a $\psi^{\prime\prime}$ charm factory~\cite{ads2,liu}
or from detailed studies of $D$ decays~\cite{aad}, this relation puts an
additional constraint on the system of equations~(\ref{eqnd}). Indeed, if
$\delta_{f_1}$ and $\delta_{f_2}$ are known then $\zeta_K$ may also be
extracted, thereby providing information about final state interaction effects
in $B$ decays. Conversely, if the left hand side of eq.~(\ref{xidelta}) is
determined from studies of CP violation, information is obtained about $D$
decay phase shifts. \par Method (2) is a straightforward generalization of the
GLW method. Instead of a CP-eigenstate, a non-CP eigenstate $f$ is used. In
addition to $Br(B^- \rightarrow K^- D^0)$, we assume that $Br(B^- \rightarrow
K^- \overline D^0)$ is accurately known as well as the following branching
ratios: \begin{eqnarray} Br(B^- \to K^- [f])\;, \;\;Br(B^+ \to K^+ [\overline
f])\; . \nonumber \end{eqnarray} \par Thus, for the mode $f$ we know $a(K)$,
$b(K)$, $c(f)$, $c(\overline f)$, $d(K,f)$ and $\overline d(K,f)$. We see that
eq.~(\ref{eqnd}) (for $f_i = f$) is now a system of two equations in two
unknowns $\{\gamma$, $\xi^K_f\}$ and can therefore be solved.  This system of
equations is identical to the geometric construction in
\cite{gronaul,gronauw,d0ks}. Using additional distinct modes $f'$ will reduce
ambiguities and determine $\gamma$ more accurately. There are several
variations and straightforward generalizations of these methods of extracting
$\gamma$, which will be discussed in detail elsewhere \cite{ads2}. \par The
discussion above as it applies to $B^-\to K^- D^0$ versus $K^- \overline D^0$
in fact may be generalized with little modification to $B$ decays of the form
$B^-\to {\bf k}^- {\bf d}^0$ versus ${\bf k}^- \overline {\bf d}^0$ where ${\bf
k}^-$ denotes $K^-$, $K^{*-}$ or any higher kaonic resonance. Likewise ${\bf
d}^0$ denotes ${D^0}$, ${D^{*0}}$ or any higher $D$-resonance where that
excited state cascades down to a ${D^0}$ that in turn decays to final states
accessible to both $D^0$ and $\overline D^0$. This immediate generalization is
constrained to cases where ${\bf k}^-$ or ${\bf d}$ is spin 0 or else several
partial waves will be present. The case with multiple partial waves may still
be considered except that each of the amplitudes may have a different strong
phase and so must be separated. Of course if this analysis can be done, it may
provide an advantage since method (1) could then be applied to several
amplitudes with the same particles in the final state. \par Let us now give a
rough numerical estimate of the typical size of the asymmetry $A(K,f)$ and the
number of $B's$ needed to observe the effect using our method. We shall perform
the estimate for the case $B^-\to K^{*-} [K^+\rho^-]$. We start with the known
branching ratio $Br(B^-\to \rho^- D^0)=1.3\%$. Multiplying this by the Cabibbo
factor of $\sin^2\theta_C$ one obtains an estimate of $a(K^*)\approx 6.6\times
10^{-4}$. Using the ratio in Eq.~\ref{ampratio}, one obtains $b(K^*)\approx
6\times 10^{-6}$. The experimental value of $c(K^-\rho^+)=.11$. To estimate the
value of $c(K^+\rho^-)$ let us suppose that
$c(K^-\pi^+):c(K^+\pi^-)=c(K^-\rho^+):c(K^+\rho^-)$, thus $c(K^+\rho^-)\approx
8.5\times 10^{-4}$. \par In terms of the angles $\xi^{K^*}_{K^+\rho^-}$ and
$\gamma$, the partial rate asymmetry $A$ is given by: \begin{eqnarray}
A(K^{*},K^+\rho^-)= - { R(K^{*},K^+\rho^-) \sin\xi^{K^*}_{K^+\rho^-}\sin\gamma
/ (1+ R(K^{*},K^+\rho^-) \cos\xi^{K^*}_{K^+\rho^-}\cos\gamma) } \label{aexprn}
\end{eqnarray} where \begin{eqnarray} R(K^*,K^+\rho^-)= {
2\sqrt{a(K^*)b(K^*)c(K^+\rho^-)c(K^-\rho^+)} \over
a(K^*)c(K^+\rho^-)+b(K^*)c(K^-\rho^+) } \label{rdef} \end{eqnarray} For the
numbers above then $R=.99$. In order to estimate the asymmetry $A$ however, we
need to know the value of the weak and strong phases which are not very well
constrained experimentally. For the purpose of our estimates, let us take
$\cos\xi^{K^*}_{K^+\rho^-}\cos\gamma=0$ so that the denominator in
eq.~(\ref{aexprn}) assumes its average value and also
$\sin\xi^{K^*}_{K^+\rho^-}\sin\gamma=1/2$ where $1/2$ is the r.m.s average
value of $\sin\theta_1\sin\theta_2$ for randomly selected
$\{\theta_1,\theta_2\}$. The resulting asymmetry is, $A\sim 50\%$. Let us now
define $N^{3\sigma}$ to be the total number of charged $B$'s (i.e.
$N^{3\sigma}=N(B^+)+N(B^-)$) required to observe the asymmetry $A$ to a
$3-\sigma$ significance. This quantity is thus given by: \begin{eqnarray}
N^{3\sigma}={18\over A^2 [d(K^*,K^+\rho^-)+\overline d(K^*,K^+\rho^-)]}
\label{upseqn} \end{eqnarray} which in this case would be $N^{3\sigma}\approx
5.9\times 10^7$. Similarly for the case of $B^-\to K^{*-} [K^+\pi^-]$
$N^{3\sigma}\approx 15\times 10^7$. \par As a comparison, one can perform a
similar estimate for the case where $f$ is a CP eigenstate as in the GLW
method. Thus if we take $f=K_S\pi^0$, and assume $\sin\zeta_K\sin\gamma=1/2$;
$\cos\zeta_K\cos\gamma=0$, we get $R\approx .19 $, $A\approx 9.5\%     $, and
finally $N^{3\sigma}\approx 31\times 10^{7}$. In the GLW method it is possible
to properly combine statistics for all CP eigenstate modes. If one does not
include modes with $K_L$ this amounts to a branching fraction which is roughly
$5\%$ of $D^0$ decays. Taking $5\%$, we find that $N^{3\sigma}\approx 6.5\times
10^{7}$, about the same as for our single mode above. In~\cite{ads2} similar
estimates are performed for the modes $B^-\to K^-[K^+\rho^-]$,
$K^{-}[K^+\pi^-]$, $K^{*-}[K^+\pi^-]$, $K^{-}[K^+a_1^-]$, $K^{*-}[K^+a_1^-]$,
$K^{-}[K^{*+}\pi^-]$ and $K^{*-}[K^{*+}\pi^-]$ each of which produces results
for $A$ and $N^{3\sigma}$ of the same order of magnitude as the $B^-\to K^{*-}
[K^+\rho^-]$ case. \par An important point to bear in mind about CP
non-eigenstate modes such as $K^{*+}\pi^-$ and $K^+\rho^-$ is that they are
just approximations to concentrations in the Dalitz plot for $K\pi\pi$. In full
generality each point of this Dalitz plot contains a separate value of
$\delta$. In principle, one can generate a set of equations~(\ref{eqnd}) at
each such point and then proceed to extract $\gamma$ as in method (1). In
practice, if the variation of the strong phase is accurately known or well
modelled, one can weight information optimally to extract $\gamma$. Such a
Dalitz plot analysis, which may be generalized to $n$-body decays, is discussed
extensively in~\cite{ads2}. Comparing such a generalized Dalitz plot of $f$ for
a $B$ decay with its CP conjugate partner could show striking CP violating
effects. The numerical estimates above do, however, provide a rough idea of the
reach of such modes. \par Finally, let us comment on $D^0$ decay modes which
are singly Cabibbo suppressed yet not CP eigenstates such as
$K^{*\pm}K^\mp,\;K^{**\pm}K^{(*)\mp},$ $\; \pi^\pm\rho^\mp,\; \pi^\pm
a_1^\mp,\rho^\pm a_1^\mp,$ etc. Since for these modes the quark content is self
conjugate, $c(f)\approx c(\overline f)$. Thus, as with the true CP eigenstate
modes of the GLW method, the CP violating effects from $B^-\to K^-[f]$ will be
$O(10\%)$ and $N^{3\sigma}$ will be similar to that estimated above for the GLW
case. On the other hand, in $B^0$ decays both the modes $B^0\to {\bf k}^0 D^0$
and $B^0\to {\bf k}^0 \overline D^0$ are color suppressed  and so
$D^0(\overline D^0)$ decays to such singly Cabibbo suppressed modes could lead
to large CP asymmetries. Indeed such an approach, which provides an additional
strong phase difference due to $D^0$ decays may significantly enhance the
methods discussed in~\cite{gronaul,d0ks} where CP eigenstates are used. \par In
summary, we reiterate the potential limitations of the GLW method:
\begin{enumerate} \begin{enumerate} \item One must observe decays of $D^0$ to a
CP eigenstate.  All such modes are either Cabibbo suppressed  or color
suppressed and the experimentally feasible total (ignoring $K_L$ modes) is less
than 5\%. \item The CP violating asymmetries from the decays of $D^0$ to CP
eigenstates are $O(10\%)$ at best, whereas more dramatic asymmetries would be
desirable. \item The GLW method requires knowledge of the branching ratios
$B(B^- \to K^- D^0)$ and $B(B^- \to K^- \overline D^0)$ where the latter may
present experimental difficulties. \item If it should happen that the strong
phase difference between ${\cal M}(B^- \to K^- D^0)$ and ${\cal M}(B^- \to K^-
\overline D^0)$ is small, then no observable CP violation will be produced even
though one may still be able to deduce $\cos\gamma$. \par \end{enumerate}
\end{enumerate} \par In our method, problem (a) is overcome since there are a
large number of different modes that one can use. Using the decay chains
$B^-\to K^- {D^0}[\to f]$ and $K^-\overline D^0[\to f]$ (where $f$ is a doubly
Cabibbo-suppressed mode of $D^0$, and thus Cabibbo favored mode of $\overline
D^0$) the event rate is reduced but this plays to our benefit since asymmetries
of $O(100\%)$ are likely in at least some modes [see problem (b)]. More
detailed estimates~\cite{ads2} show that the required number of $B$ events are
favorable (at worst comparable) in comparison to the original GLW method for
extracting the CKM angle $\gamma$. Problem (c) can be circumvented because we
can dispense with the need to know $Br(B^- \to K^- \overline D^0)$ by
considering different hadronic final states  $f_i$ of neutral $D$ mesons with
different strong phases. In such cases we can solve for $Br(B^- \to K^-
\overline D^0) / Br(B^- \to K^- D^0)$ and $\gamma$. Problem (d) is unlikely in
our case because for non-CP eigenstate modes $f_i$, the strong phase difference
between the two interfering $B$ decay amplitudes [${\cal M}(B^- \to K^- D^0)$
and  ${\cal M}(B^- \to K^- \overline D^0)$] is combined with an additional
strong phase difference in $D$ decays, $\delta_{f_i}$. Judicious choices of
$D^0$ modes thus allow potentially large strong phase differences, thereby
significantly enhancing CP violating effects in $B$ decays involving
$D^0/\overline D^0$ mesons. \par In closing, we recall that various $B$
detectors currently under construction are specifically designed to observe
mixing-induced CP violation. Such experiments should be able to determine the
CKM phase $\beta$ without any assumption concerning strong phases. Likewise
both for the original GLW method~\cite{gronauw} and our version, $\gamma$ is
reconstructed (up to discrete ambiguities) without any assumption about the
value of the strong phase. The ability to probe $\gamma$ more incisively,
improves our capacity to constrain or rule out the standard model. In addition,
since these methods measure direct CP violation rather than oscillation
effects, one may perform such experiments at any facility where $B$ mesons are
copiously produced. Because neither tagging nor time-dependent studies are
required, such effects could be observed at even a symmetric $\Upsilon(4S)$
factory, such as CLEO. To optimize the observation and interpretation of such
effects, accurate measurements of the relevant $D^0$ decays are highly
desirable. \par \bigskip \par \bigskip \par We thank J.~Butler and P.~Lebrun
for stressing the significance of final state interaction effects in $D^0$
decays. I.D. thanks R.G. Sachs for emphasizing the importance of non-CP
eigenstates in studies of CP violation and R. Aleksan, B. Kayser, and F. Le
Diberder for pleasant collaborations on related issues. This research was
supported in part by the U.S. DOE contracts DC-AC05-84ER40150 (CEBAF),
DE-AC-76CH00016 (BNL) and DE-AC02-76CH03000 (FNAL). \par  
\begin{references}
\bibitem{suppressed} A more refined estimate must include the color suppressed
amplitude present for the $B^- \to K^- D^0$ process. \par
\bibitem{foot} In our discussion, we
neglect $D^0 -\overline D^0$ mixing and CP violation in $D^0$ decays, since the
Standard Model predicts them to be orders of magnitude smaller than the effects
highlighted in this note. \par
\bibitem{d0bartag} The signal via semileptonic $\overline D^0$ decays does have
some distinguishing features which may reduce backgrounds. First, in the $B^-$
frame the $K^-$ is monochromatic. Second, the signal involves two kaons.
Finally, the semileptonic decay of the $\overline D^0$ originates from a
tertiary vertex, in contrast to background events. \par
\end{references}
\end{document}